\documentclass[mathleft,fleqn,%
]{an}
\usepackage{graphicx}
\usepackage[varg]{txfonts}
\usepackage{natbib}
\bibpunct{(}{)}{;}{a}{}{,}
\newcommand{\marque}[1]{\textbf{{[#1]}}}
\newcommand{\ind}[1]{_{\mathrm{#1}}}
\newcommand{\diff}{\mathrm{d}}

\def\Kepler{\emph{Kepler}}

\def\numax{\nu\ind{max}}\def\nmax{n\ind{max}}
\newcommand{\np}{n}

\def\d0l{d_{0\ell}}
\def\Dnu{\Delta\nu}

\def\Dnuobs{\Delta\nu\ind{obs}}
\def\Dnuas{\Delta\nu\ind{as}}

\def\Tg{\Delta\Pi_1}
\def\epsobs{\varepsilon\ind{obs}}

\def\Dnuas{\Delta\nu\ind{as}}
\def\epsas{\varepsilon\ind{as}}

\newcommand\dnurot{\delta\nu\ind{rot}}

\newcommand\thetap{\theta\ind{p}}
\newcommand\thetag{\theta\ind{g}}
\newcommand\epsp{\varepsilon\ind{p}}

\newcommand\nup{\nu\ind{p}}\newcommand\nug{\nu\ind{g}}
\newcommand\Dnup{\Delta\nu_n}

\def\Amax{A\ind{max}}

\newcommand{\BV}{Brunt-V\"ais\"al\"a}

\newcommand\iref{_\odot}

\newcommand\dnusas{\Dnu\ind{ref}}
\newcommand\numaxas{\nu\ind{ref}}
\newcommand\Teff{{T\ind{eff}}}
\newcommand\Ts{T\iref}
\newcommand\Rs{R\iref}
\newcommand\Ms{M\iref}
\newcommand\Rsis{R} %\newcommand\Rsis{R\ind{s}}
\newcommand\Msis{M} %\newcommand\Msis{M\ind{s}}

\begin{document}

% The following seven commands are intended for editorial usage and
% should be ignored by the author(s).
\Pagespan{1}{}% Document's page range.
% If second parameter is left empty, the last page is computed
% automatically.
\Yearpublication{2015}%
\Yearsubmission{2015}%
\Month{0}%
\Volume{999}%
\Issue{0}%
\DOI{asna.201400000}%

\title{Seismic indices -- a deep look inside evolved stars}

\author{B. Mosser\inst{1}\fnmsep\thanks{Corresponding author:
        {benoit.mosser@obspm.fr}}
}
\titlerunning{Seismic indices}
\authorrunning{B. Mosser}
\institute{ LESIA, Observatoire de Paris, PSL Research University,
CNRS, Universit\'e Pierre et Marie Curie, Universit\'e Paris
Diderot,  92195 Meudon, France}

\received{XXXX}
\accepted{XXXX}
\publonline{XXXX}


\abstract{%
Independent of stellar modelling, global seismic parameters of red
giants provide unique information on the individual stellar
properties as well as on stellar evolution. They allow us to
measure key stellar parameters, such as the stellar mass and
radius, or to derive the distance of field stars. Furthermore,
oscillations with a mixed character directly probe the physical
conditions in the stellar core. Here, we explain how very precise
seismic indices are obtained, and how they can be used for
monitoring stellar evolution and performing Galactic archeology.}

\maketitle

%--------------------------------------------------------------
\section{Introduction}

Red giant seismology is one of the exquisite surprise provided by
the space missions CoRoT and \Kepler. From, the analysis of long,
continuous, ultra-precise photometric light curves, seismic
indices can be derived from calibrated global seismic parameters
that describe the oscillation pattern. They measure the properties
of both the stellar envelope  and the core. In this review, we
focus on ensemble asteroseismology results that provide a wealth
of global information. The analysis and modelling of individual
stars have started for a handful of targets
\citep[e.g.,][]{2011MNRAS.415.3783D,2011ApJ...742..120J,2012A&A...538A..73B,2014A&A...562A.109L,2015arXiv151106160D}
but are not presented here, despite the fact they are crucial for
the deep understanding of the stellar interior structure and of
the physical input to be considered, such as the measurement of
the location of the helium second-ionization region
\citep{2010A&A...520L...6M}, or the measurement of differential
rotation \citep{2012Natur.481...55B,2012ApJ...756...19D}. The
global properties of the low-degree oscillation spectra used to
derive relevant estimates of the stellar masses and radii are
presented in Section \ref{radial}. In Section \ref{dipole}, I show
the rich information provided by the identification of mixed
modes. They result from the coupling of gravity waves propagating
in the radiative core region, with pressure waves mainly
propagating in the stellar envelope, and directly reveal
information from the stellar core \citep{2011Natur.471..608B}. A
few results permitted by ensemble asteroseismic measurements are
presented in Section \ref{indices}.

%--------------------------------------------------------------
\section{Solar-like oscillations\label{radial}}

Two seismic indices, the frequency $\numax$ of maximum oscillation
signal and the frequency separation $\Dnu$ between radial modes,
can be easily defined from the structure of the low-degree
oscillation pressure pattern. In this Section, we intend to show
how they can be precisely determined.

\subsection{Gravity and convection}

The empirical definition of $\numax$ is clear. However, measuring
it is more difficult,  partly because $\numax$ depends on the
observable (intensity or velocity measurements provide two
different values of $\numax$ since the translation of relative
photometric amplitude to Doppler velocity depends on frequency).
In fact, we lack a precise and non-empirical definition of
$\numax$. In practice, the maximum of oscillation signal comes
from the minimum of the product $\eta \,I$, where $\eta$ is the
damping rate and $I$ the inertia of the modes.
\citep[e.g.,][]{1992MNRAS.255..603B,2012A&A...540L...7B}. This
occurs when oscillation periods are equivalent to the thermal and
convective time scales.

Observations show that $\numax$ provides a highly-precise
measurement of the stellar gravity
\citep[e.g.,][]{2012MNRAS.419L..34M,2014A&A...564A.119M,2014ApJS..215...19P}
since it scales as the acoustic cutoff frequency
\citep{1991ApJ...368..599B,2011A&A...530A.142B}, hence as the
ratio $g/\sqrt{\Teff}$. This is verified for most of stars, except
in some cases, as Procyon, where the oscillation excess power
shows two humps \citep{2008ApJ...687.1180A}. In the general case,
from $\numax$ we get a reliable and precise proxy of $M \,
R^{-2}\, \Teff^{-1/2}$.

\subsection{Homology and universal pattern}

The low-degree oscillation spectrum of red giant pressure modes
follows, as for other stars, the asymptotic expansion
\citep{1980ApJS...43..469T}:
\begin{equation}\label{eqt-asymp}
    \nu_{n,\ell} = \left( n'_\ell + {A_\ell \over n'_\ell}\right) \
    \Dnuas \hbox{ ,\ with \ } n'_\ell = n  + {\ell\over 2} + \epsas
    ,
\end{equation}
where $n$ is the radial order and $\ell$ is the angular degree.
The asymptotic large separation $\Dnuas = (2\int_0^R \diff r
/c)^{-1}$ measures the stellar acoustic radius. The asymptotic
value of $ \epsas$ is 1/4. The second-order coefficients $A_\ell$
have a more complicate form.

CoRoT observations have evidenced a unique property of the red
giant oscillation pattern: following the interior structure
homology, the pattern can also be defined as homologous. The
concept of universal red giant oscillation pattern was  introduced
by \cite{2011A&A...525L...9M}, as an alternative form to the usual
asymptotic expansion \citep{1980ApJS...43..469T}, with the
observed large separation as the only free parameter. The
second-order asymptotic expansion expresses as
\begin{equation}\label{ordre_deux}
    \nu_{n,\ell} = \left( n+{\ell\over 2} +\epsobs + \d0l + {\alpha \over 2}\, (n-\nmax)^2\right) \
    \Dnuobs ,
\end{equation}
where all the parameters depend either on $\Dnuobs$ or on the
dimensionless parameter $\nmax=\numax/\Dnuobs - \epsobs$
\citep{2013A&A...550A.126M}: the radial offset $\epsobs$ helps to
locate the radial ridge; the non-radial offsets $\d0l$ express the
shifts of the different degrees $\ell$ compared to the radial
modes \citep[e.g.,][]{2012ApJ...757..190C}; the term $\alpha$
accounts for the second-order asymptotic expansion.

\begin{figure}[t]
\includegraphics[width=8.cm]{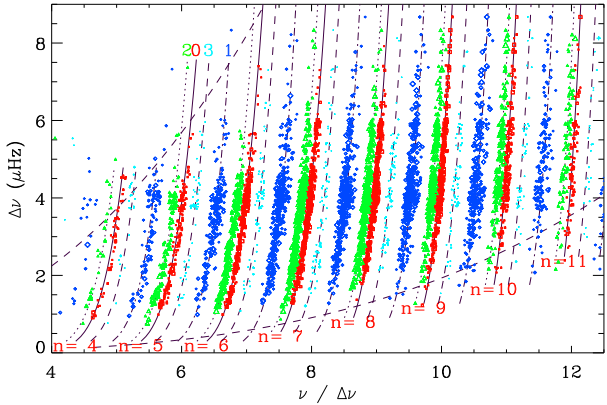}
\caption{Identification of the low-degree oscillation modes of
CoRoT red giants. Each color corresponds to a different mode
degree (radial modes in red, dipole modes in dark blue, $\ell=2$
modes in green, $\ell=3$ modes in light blue). The solid grey
lines indicate the fits of $\varepsilon$ for each radial order
$n$. The fits of $d_{01}$, $d_{02}$ and $d_{03}$ are superimposed
on the respective ridges (respectively dash-dot, dot, and dash
lines for $\ell = $1, 2, and 3). The dark dashed lines delineate
the region where the modes have noticeable amplitudes. Figure from
\cite{2011A&A...525L...9M}. \label{fig-identi}}
\end{figure}

The signature of homology of red giants is illustrated in
Fig.~\ref{fig-identi}, where CoRoT red giant oscillation spectra
sorted by increasing large separation values are plotted on the
same graph, with a dimensionless frequency in abscissa. The
alignment of the ridges, each one corresponding to a given radial
order $n$ and angular degree $\ell$, demonstrates the validity of
Eq.~(\ref{ordre_deux}). Contrarily, oscillation patterns in
main-sequence stars could not be superposed.

Equation~(\ref{ordre_deux}) accounts for the measurement of
$\Dnuobs$ around $\numax$. The asymptotic large separation
$\Dnuas$ corresponds to the frequency spacing at very high
frequency, hence it cannot be measured. Its theoretical value is
slightly larger than the observed value; the relationship between
the observed and asymptotic parameters is studied in
\cite{2013A&A...550A.126M}. The high accuracy level reached by
Eq.~(\ref{ordre_deux}) has been shown in previous comparison work
\citep[e.g.,][]{2011MNRAS.415.3539V,2012A&A...544A..90H}; the
accuracy of the measurement of $\Dnuobs$ is better than
$0.04\,\mu$Hz for all evolutionary stages
\citep{2013EAS....63..137M}.

The measurement of $\Dnuobs$ is however perturbed by a modulation
due to the rapid local variation of the sound speed in the stellar
interior, related to the density contrast at the core boundary or
to the local depression of the sound speed that occurs in the
helium second-ionization region \citep{2010A&A...520L...6M}.  The
main signature of the glitch induces a modulation of the spectrum.
Ensemble analysis of this glitch has shown how it can be modelled,
depending on stellar evolution \citep{2015A&A...579A..84V}. This
means that we can derive glitch-free measurement of the large
separation. As a result, with the large separation measured in a
broad frequency range, accounting for the second-order correction
of the asymptotic correction, and corrected for glitch
perturbation, the measurement of the large separation provides a
reliable proxy of the square root of the mean density $\sqrt{M /
R^3}$.

\begin{figure*}
\includegraphics[width=6.7cm]{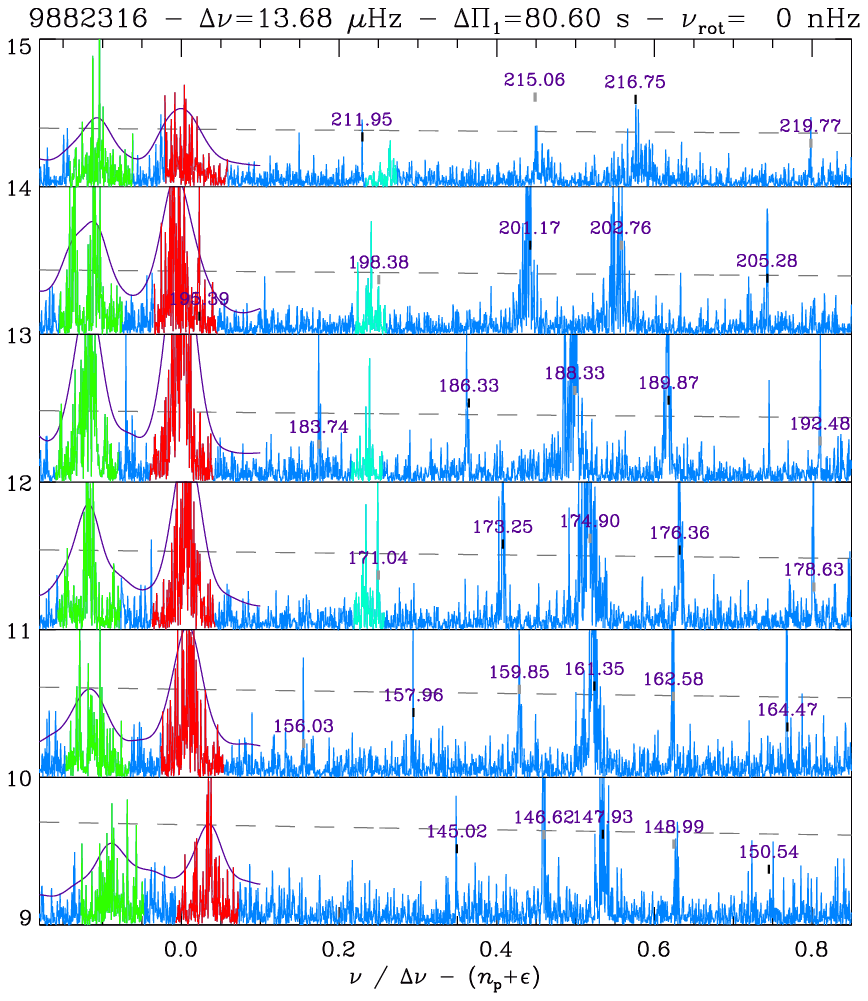}
\includegraphics[width=4.4cm]{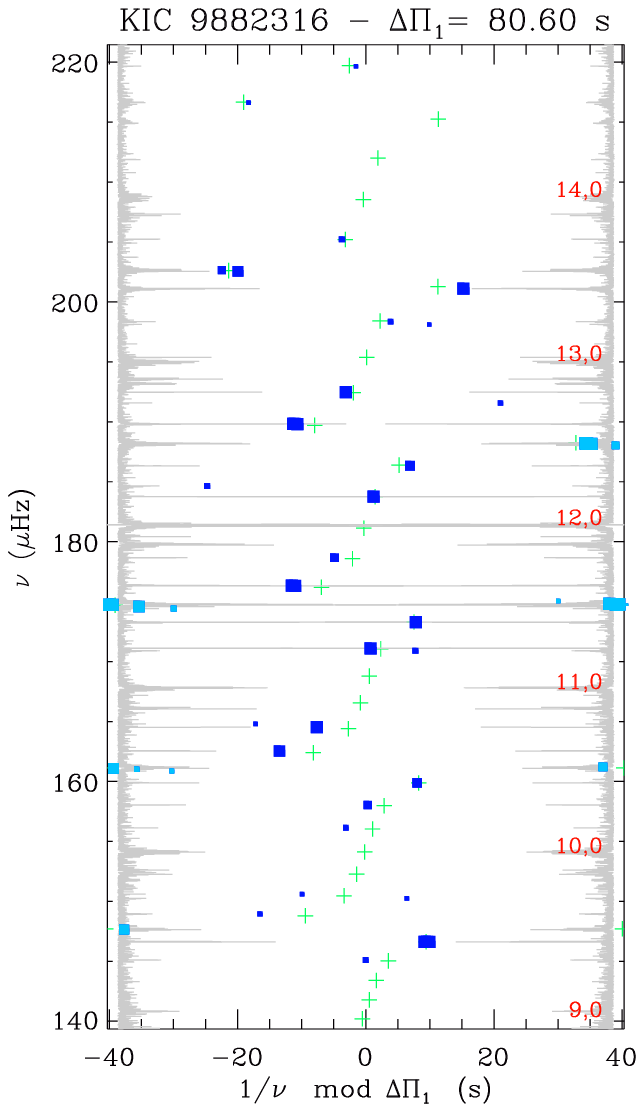}
\includegraphics[width=4.4cm]{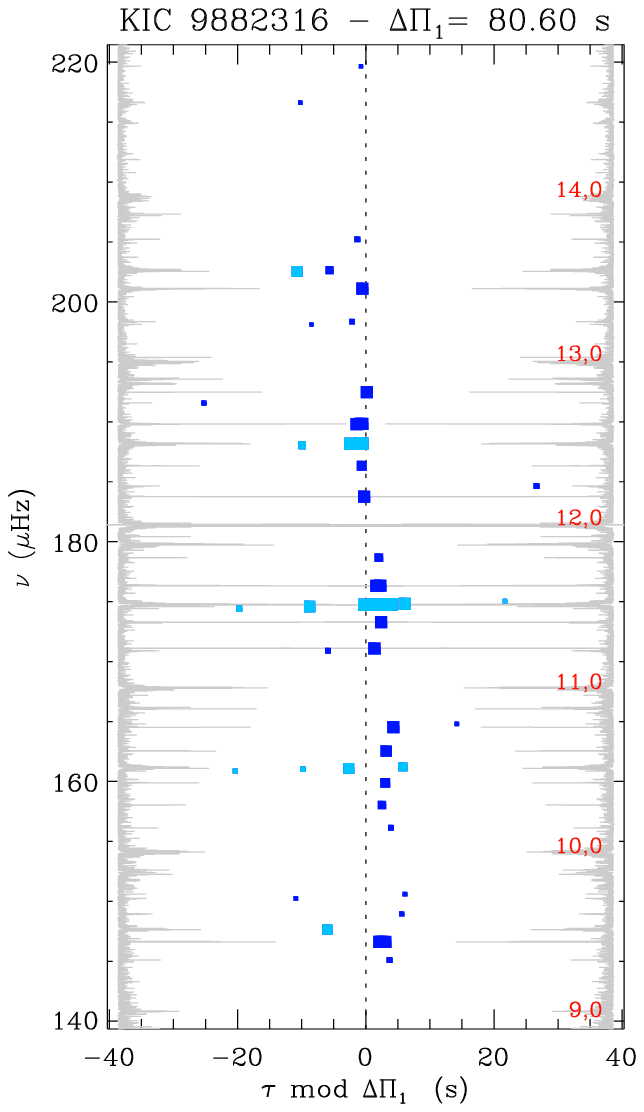}
\caption{\'Echelle diagrams of the RGB star KIC 9882316 that shows
no rotational signature.
 {\textsl{Left}:} Frequency \'echelle diagram  as a
function of  $\nu/\Dnu - (\np+\epsp)$. The radial order $\np$ is
indicated on the y-axis. Radial modes are highlighted in red,
$\ell=2$ modes in green, and $\ell=3$ modes, when observed, in
light blue. Dipole mixed mode frequencies with a height larger
than eight times the mean background value (grey dashed lines) are
identified in $\mu$Hz.
 {\textsl{Middle}:} Period \'echelle diagram, based on $1/\nu$ modulo $\Tg$. The most prominent mixed
modes, marked with blue filled squares (in light blue for peaks in
the vicinity of pure pressure modes), are automatically
identified. In the background of the figure, the spectra are
plotted twice and top to tail for making the mode identification
easier, with $\np$ indicated for radial modes.
{\textsl{Right}:} Stretched period \'echelle diagram, based on
$\tau$ modulo $\Tg$. Figure from \cite{2015A&A...584A..50M}
}\label{fig-echelle}
\end{figure*}

\subsection{Scaling relations}

Seismic relations with $\Dnuobs$ and $\numax$ can be used to
provide proxies of the stellar masses and radii. The scaling
relations write
\begin{equation}
  {\Rsis \over\Rs}  = \left({\numax \over \numaxas}\right) \
     \left({\Dnuas \over \dnusas}\right)^{-2}
     \left({\Teff \over \Ts}\right)^{1/2}, \label{scalingRas}
\end{equation}
\begin{equation}
  {\Msis\over\Ms} = \left({\numax \over \numaxas}\right)^{3}
     \left({\Dnuas \over \dnusas}\right)^{-4} \left({\Teff \over \Ts}\right)^{3/2} , \label{scalingMas}
\end{equation}
where the calibrated references $\numaxas = 3104\,\mu$Hz and
$\dnusas = 138.8\,\mu$Hz are derived from the comparison with
modelling \citep{2013A&A...550A.126M}, in contrast to many works
that use the solar case as a reference
\citep[e.g.,][]{2011ApJ...738L..28V,2011ApJ...740L...2S,2013ApJ...765L..41S}.
Large efforts are undertaken for calibrating
Eqs.~(\ref{scalingRas}) and (\ref{scalingMas}), since solar
reference values provide biases in the red giant regime
\citep[e.g.,][]{2011ApJ...743..161W,2012MNRAS.419.2077M,2014ApJ...785L..28E},
as made clear since CoRoT observations
\citep{2010A&A...509A..77K,2010A&A...517A..22M}. Following
\cite{2013ASPC..479...61B}, we note that, for performing the
calibration, it is necessary to avoid confusion between the large
separation $\Dnuobs$ measured from radial modes observed around
$\numax$, the asymptotic large separation $\Dnuas$, and the
dynamical frequency $\nu_0$ that scales with $\sqrt{M/R^3}$.

%--------------------------------------------------------------
\section{Dipole mixed modes\label{dipole}}

Mixed modes allow us to detect the gravity modes that probe the
stellar core in a much more efficient way compared to the Sun
\citep[e.g.,][]{2009A&A...494..191B}.

\subsection{Asymptotic expansion}

\cite{1979PASJ...31...87S} and \cite{1989nos..book.....U} derived
an implicit asymptotic relation for mixed modes in case of weak
coupling, which expresses as
\begin{equation}\label{eqt-mixed}
    \tan\thetap = q \tan\thetag .
\end{equation}
The phases $\thetap$ and $\thetag$ refer, respectively, to the
pressure- and gravity-wave contributions, and $q$ is the coupling
factor between them. \cite{2015A&A...584A..50M} write them
\begin{equation}
   \thetap =  \pi {\nu-\nup\over \Dnup}
   \ \hbox{ and } \
   \thetag = \pi {1 \over \Tg}  \left({\displaystyle{1\over\nu}
  -\displaystyle{1\over\nug}}\right)
   \label{eqt-pg}
  ,
\end{equation}
where $\nup$ and  $\nug$ are the asymptotic frequencies of pure
pressure and gravity modes, respectively, $\Dnup$ is the frequency
difference between two consecutive pure pressure radial modes (the
subscript $n$ means that small variations of the large separation
with the radial order can be accounted for), and $\Tg$ is the
asymptotic period spacing of gravity modes.
Equation~(\ref{eqt-pg}) can be used at any order of the asymptotic
expansions for the pure pressure and gravity contributions. For
instance, these expressions may include the signature of acoustic
or buoyancy glitches \citep{2015A&A...584A..50M}.

\subsection{Period spacing and rotational splitting}

The derivation of the implicit asymptotic expansion provides an
analytical form of the period spacing between two consecutive
mixed modes, which writes ${\Delta P / \Tg} = \zeta$, with
\begin{equation}\label{eqt-zeta}
    \zeta = \left[1+ {1\over q}
     {\nu^2 \Tg \over \Dnup}
    {\cos^2 \pi \displaystyle{1\over \Tg}
    \left(\displaystyle{{1\over \nu} - {1\over\nug}}\right)
    \over
    \cos^2 \pi \displaystyle{\nu-\nup\over \Dnup}}
    \right]^{-1} .
\end{equation}
Interestingly, this expression matches the first-order asymptotic
expansion derived by \cite{2015A&A...580A..96D} for expressing the
mixed-mode rotational splitting. Following
\cite{2013A&A...549A..75G} and \cite{2015A&A...580A..96D}, the
rotational splitting of mixed modes expresses
\begin{equation}\label{eqt-zeta-rotations}
     \dnurot =
     {\dnurot}\ind{,g}\ \zeta
     +
     {\dnurot}\ind{,p}\ (1-\zeta)
     ,
\end{equation}
where ${\dnurot}\ind{,g}$ and ${\dnurot}\ind{,p}$ are the
rotational splittings related to pure gravity or pure pressure
modes. The measurement of rotational splittings and the derivation
of the mean core rotation are discussed in
\cite{2012A&A...548A..10M}. Specific cases are discussed in
\cite{2012Natur.481...55B}, \cite{2012ApJ...756...19D}, and
\cite{2014A&A...564A..27D}.

\subsection{Asymptotic period spacing}

The previous developments provide the theoretical basis for the
thorough understanding of the mixed modes. Following
\cite{2015A&A...584A..50M}, the function $\zeta$ is used to turn
the frequencies into periods $\tau$, with
%\begin{equation}\label{eqt-stretch}
%    \diff\tau = {1\over \zeta} {\diff \nu \over \nu^2} .
%\end{equation}
$  \diff\tau = -\diff\log(1/\nu)/ \zeta$. From the integration of
this change of variable we derive corrected periods $\tau$ of
mixed modes, called stretched periods. \'Echelle diagrams based on
$\tau$ show the structure of the gravity modes contributing to the
mixed modes (Fig.~\ref{fig-echelle}). This allows the most
accurate measurement of the asymptotic period spacing even when
rotation splittings are important \citep{vrard} and emphasizes the
signature of the buoyancy glitches, namely rapid variation of the
\BV\ frequency in the radiative core.

\begin{figure}[t]
 \centering
 \includegraphics[width=7.8cm]{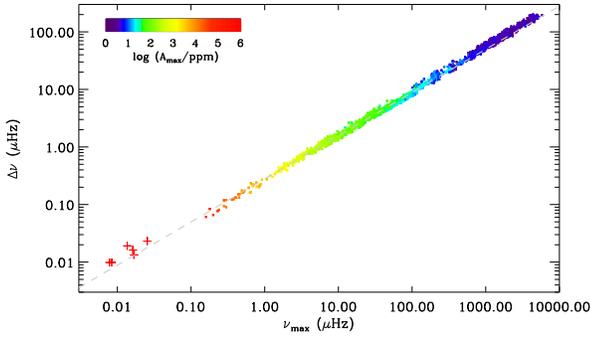}
%% Note the ABSENCE of the extension .pdf , .eps or .ps  !
  \caption{$\numax$ -- $\Dnuobs$ relation, with CoRoT and \Kepler\ data.
  Plusses are red supergiants showing observed by
  \cite{2006MNRAS.372.1721K}, with $\Dnuobs$ and $\numax$
  derived from a combination of the frequencies.
  The colors code the mean maximum amplitude $\Amax$ of the radial oscillations.
  The grey dashed line has a slope 3/4.}
  \label{numaxdnu}
\end{figure}

\begin{figure}[t]
 \centering
 \includegraphics[width=7.2cm]{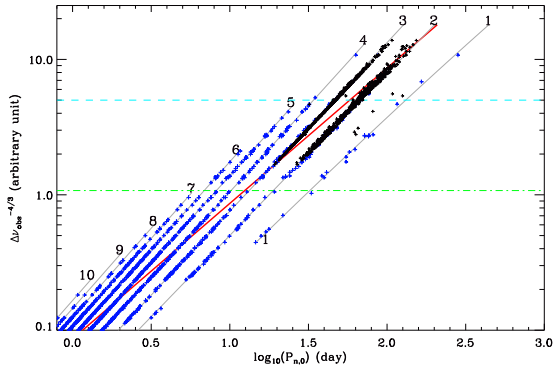}
%% Note the ABSENCE of the extension .pdf , .eps or .ps  !
  \caption{Period-luminosity relations of radial modes of AGB and RGB stars
observed with \Kepler\ (blue pluses) and OGLE (black symbols). The
proxy of the luminosity is derived from $\Dnuobs^{-4/3}$. Each
sequence, corresponding to a fixed radial order, is fitted with
the model provided by Eq.~(\ref{ordre_deux}) extrapolated to low
frequency (gray solid lines). The thick red line indicates the
location of $\numax$. The blue dashed line corresponds to the tip
of the RGB. Figure from \cite{2013A&A...559A.137M}.}
  \label{fig-PL}
\end{figure}

%--------------------------------------------------------------
\section{Seismic indices\label{indices}}

Global seismic parameters can be used as seismic indices for
characterizing stars and performing ensemble asteroseimology.

\subsection{Stellar evolution up to the AGB\label{ogle}}

From the scaling relations Eqs.~(\ref{scalingRas}) and
(\ref{scalingMas}), we derive
\begin{equation}\label{scaling-dnu}
   \Dnuobs  \simeq \Dnuas \propto M^{-1/4} \ \Teff^{3/8} \
   \numax^{3/4}
   .
\end{equation}
On the red and asymptotic giant branches, all low-mass stars are
present at all evolutionary stages, so that the mass play no role
in this relation. Consequently, $\Dnuobs$ scales as $\numax^{3/4}$
(Fig.~\ref{numaxdnu}). The validity of Eq.~(\ref{scaling-dnu})
over more than six decades in frequency indicates that the stellar
red giant populations observed by CoRoT or \Kepler\ constitute a
set of stars homogenous enough to mimic stellar evolution.

This result extrapolated to very low $\numax$, with
Eq.~(\ref{ordre_deux}) iteratively adapted to  fit the
low-frequency oscillation spectra, was used for demonstrating that
period-luminosity relations in semi-regular variables are drawn by
low-degree low-radial-order solar-like oscillations
\citep{2013A&A...559A.137M}. When the large separation decreases,
the radial orders of the observed modes decrease too, down to
$\nmax=2$ (Fig.~\ref{fig-PL}). Interpreting oscillations in
semi-regular variables as solar-like oscillations can be used to
investigate with a firm physical basis the time series obtained
from ground-based microlensing surveys. This will provide improved
distance measurements, since an accurate measurement of the
stellar radius permits an accurate use of the Stefan-Boltzmann
law, and opens the way to extragalactic asteroseismology, with the
observations of M giants in the Magellanic Clouds.
\cite{2013A&A...559A.137M} have also shown that the acceleration
of the external layers of red giant with solar-like oscillations
is about the same order of magnitude as the surface gravity when
the stars reach the tip of the RGB: global oscillations play a
role in the mass-loss process.

\begin{figure}[t]
  \includegraphics[width=7.3cm]{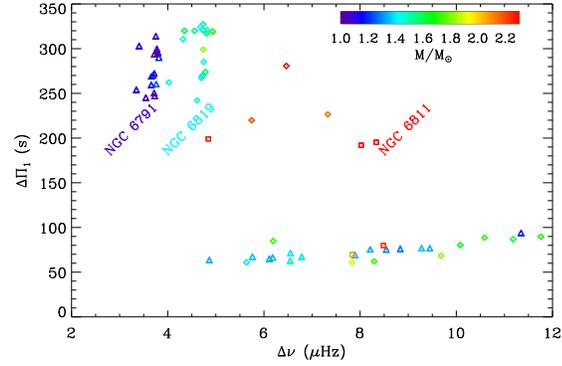}
  \caption{$\Tg$ -- $\Dnu$ relations
  for red giants in three open clusters observed by
  \Kepler (NGC 6791, $\small{\triangle}$; 6819, {\large{$\diamond$}}, with three blue stragglers; 6811,
  {\small{$\square$}}. Tracks on the RGB are superimposed, but not in the
  red clump.
  Figure from \cite{vrard}.
  }
  \label{fig:clusters}
\end{figure}

\subsection{$\Dnu$ -- $\Tg$ diagram}

The measurement of $\Tg$ provides a unique view on the physical
properties in the stellar core \citep{2012A&A...540A.143M} and
reveals clear  differences between evolutionary stages
\citep{2014A&A...572L...5M}. The analysis, now expanded to large
populations by \cite{vrard} with about 5\,000 red giants, shows
the influence of the stellar mass and metallicity on the evolution
on the RGB in the $\Dnu$ -- $\Tg$ diagram. In
Fig.~\ref{fig:clusters}, we illustrate the mass dependence of the
$\Dnu$ -- $\Tg$ relation for core-helium burning stars in three
open clusters \citep{2010ApJ...713L.182S,2012MNRAS.419.2077M}.
Evolution tracks largely depend on the mass, with massive stars in
NGC 6811 in the secondary clump. Three stars of NGC 6819, much
more massive than expected, are confirmed as blue stragglers
\citep{2012ApJ...757..190C}. The highly precise evolutionary
tracks in the $\Dnu$ -- $\Tg$ diagram can now be used to constrain
stellar evolution modelling
\citep[e.g.,][]{2015MNRAS.453.2290B,2015arXiv151203656L}.

%--------------------------------------------------------------
\subsection{Rotation}

Rotation also benefitted from ensemble asteroseismology.
\cite{2012A&A...548A..10M} have shown a significant spin-down of
the core rotation with evolution on the RGB. Mechanisms able to
efficiently transfer angular momentum from the core to the
envelope have been investigated
\citep[e.g.,][]{2012A&A...544L...4E}; the role of mixed modes for
spinning down the core rotation is underlined by
\cite{2015A&A...579A..30B,2015A&A...579A..31B}. This topic, as all
others, now benefit from supplementary non-seismic measurements
and from theoretical developments, to make the best of seismic
indices.

\acknowledgements We acknowledge the CoRoT and \Kepler\ teams,
whose efforts made these results possible. We acknowledge
financial support from the Programme National de Physique
Stellaire (CNRS/INSU) and from the ANR program IDEE.
%Interaction Des \'Etoiles et des Exoplan\`etes.

% Example of using BiBTeX (plus natbib):
% For details see \cite{1999MNRAS.309..731B},
% \cite{1893PASP....5..204C},
% \cite{2008IAUS..252...75L}. It has been demonstrated that this
% is important \citep{2012AN....333..663S}.

% Use this code if you wish to generate your bibliography with BibTeX;
% please replace first the string "an-demo" below with the name(s) of
% the BibTeX data base(s) you want to use.
% The resulting bibliography-output (the contents of the .bbl file)
% must be pasted into this file before submission.
%
%\bibliographystyle{an}
%\bibliography{an-demo}
%
% Replace the following example bibliography with your references
% before submission:

\bibliographystyle{an}
\bibliography{biblio_rg}

\appendix

\end{document}